# Local structure maturation in high entropy oxide (Mg,Co,Ni,Cu,Zn)$_{1-x}$(Cr,Mn)$_x$O thin films

Gabriela E. Niculescu[1,‡], Gerald R. Bejger[1,2,‡], John P. Barber[1,2,‡], Joshua T. Wright[3], Saeed S. I. Almishal[4], Matthew Webb[5], Sai Venkata Gayathri Ayyagari[4], Jon-Paul Maria[4], Nasim Alem[4], John T. Heron[5], Christina M. Rost[1,2,*]

1. Department of Physics and Astronomy, James Madison University, Harrisonburg, VA 22807, USA.

2. Department of Materials Science and Engineering, Virginia Polytechnic Institute and State University, Blacksburg, VA 24061, USA.

3. Department of Physics, Illinois Institute of Technology, Chicago, IL 60616, USA.

4. Department of Materials Science and Engineering, The Pennsylvania State University, University Park, PA 16802, USA.

5. Department of Materials Science and Engineering, University of Michigan, Ann Arbor, MI 48104, USA.

   *corresponding author: cmrost@vt.edu

   ‡Authors contributed equally to this work.

**Highlights:**

- Investigate novel 7-component HEO composition (MgNiCuCoZn)$_{0.167}$(MnCr)$_{0.083}$O.

- Increased pO2 pressure causes lattice parameter decrease until phase separation.

- Cr, Co, and Mn valence states highly responsive to deposition pressure.

- TEM shows a rock salt structure at low pO$_2$, shifts to spinel structure at high pO$_2$.

- Local structures shift from 6-fold to 4-fold coordination with increased oxygen.

**Abstract:**

High entropy oxides (HEO)s have garnered much interest due to their available high degree of tunability. Here, we study the local structure of (MgNiCuCoZn)$_{0.167}$(MnCr)$_{0.083}$O, a composition based on the parent HEO $(MgNiCuCoZn)_{0.2}O$. We synthesized a series of thin films via pulsed laser deposition at incremental oxygen partial pressures. X-ray diffraction shows lattice parameter to decrease with increased $pO_2$ pressures until the onset of phase separation. X-ray absorption fine structure shows that specific atomic species in the composition dictate the global structure of the material as Cr, Co, and Mn shift to energetically favorable coordination with increasing pressure. Transmission electron microscopy analysis on a lower-pressure sample exhibits a rock salt structure, but the higher-pressure sample reveals reflections reminiscent of the spinel structure. In all, these findings give a more complete picture on how $(MgNiCuCoZn)_{0.167}(MnCr)_{0.083}O$ forms with varying initial conditions and advances fundamental knowledge of cation behavior in high entropy oxides.

**Keywords:**

high entropy oxides (HEOs), X-ray absorption fine structure (XAFS), pulsed laser deposition (PLD), thin films

## 1. Introduction:

The emergence of high entropy oxides (HEOs) in materials science has opened the possibilities of novel ceramics design with a high tunability space. In the simplest definition, HEOs are usually composed of four or five cations present in equimolar or near equimolar amounts that randomly occupy relevant sites over the crystalline sublattice [1]. The seminal HEO was synthesized in 2015 of the structure $(MgZnNiCoCu)_{0.2}O$ and commonly referred to as J14 [2]. Since then, HEOs have garnered much attention for their wide range of compositions and potential use in electrical, magnetic, and optical fields, among other applications [3], [4], [5], [6], [7], [8], [9], [10]. J14 consists of a single phase, disordered oxide exhibiting the rock salt structure and containing an equal distribution of Mg, Co, Ni, Cu, and Zn on the cation sublattice. The possible compositions, structures, and applications of HEOs and related ceramics are wide ranging and continue to emerge and expand [11], [12], [13], [14], [15], [16], [17], [18].

The seminal HEO exhibited a rock salt structure, but since then, several other crystal symmetries have been successfully synthesized using a variety of techniques including solid-state method [2], [5], pyrolysis [19], floating zone [20], and pulsed laser deposition [21], [22], [23], [24]. Spinel [7], [8], [14], [25], [26] and perovskite-type structures [5], [27], [28] have gained much attention in recent years due to their more complex sublattices, which can be utilized for unique property tuning [29]. Initial studies explored well known arguments, such as the Goldschmidt tolerance factor [30], to explain the observed stabilities of perovskites. Notably, spinel-based compositions enable cations normally prone to lower-symmetry structures to be incorporated into a higher order symmetry system. HEO spinels have also been shown to have a smaller band gap than any of the parent oxides in the composition, expanding the repertoire of synergistic effects brought about by the high entropy landscape[8]. Alkali metal-doped HEO compositions were measured to have a large dielectric constant and superior ionic conductivity enabling applications in energy storage [3], [4].  High entropy rare-earth oxides (HREOs) added fluorite and bixbyite structures to the list [31], [32]. HREOs are marked by a low thermal conductivity and a tunable band gap that can be used in coatings and optics, respectively.[33], [34], [35]   Other properties currently under investigation show promise that HEOs may also find applications in ferroelectrics and magnetism [14], [28], [36], [37], [38], [39].

Much prior research has focused on the applications and properties of these HEOs as well as characterizing the crystal structure of the compositions. However, less research has been done regarding local structure trends, which has profound effects on observed properties. For example, the octahedral rotation distortions in perovskite oxides have resulted in the emergence of two-dimensional ferroelectricity [40]. Mathematically, the ideal HEO solid solution would have random occupancy, with each permutation of cation arrangement being energetically equivalent. However, Rost, et al. demonstrated that the $Cu^{2+}$ cation exhibits Jahn-Teller distortions [41] within the J14 structure, using extended X-ray absorption fine structure spectroscopy (EXAFS). These distortions show that different cations behave differently despite the global structure being the same. More recent works exploring local structure have shown varying degrees of disorder among cations [7], [13], [14], [42]. Fundamentally, this means that while a system may be high entropy, other factors such as enthalpy cannot be ignored [43], [44], which is a complex and ongoing discussion amongst the community.

In this work, we investigate the cation valence and local structure behavior of select representative atomic species in a series of high entropy oxide thin films grown using pulsed laser deposition under varying $pO_2$ to advance the fundamental understanding of such materials behavior from a local perspective. In doing so, we also test the efficacy of typical analysis methods of X-ray absorption near-edge fine structure (XANES) commonly used as applied to HEOs. Further, by understanding the underlying behaviors that govern synthesis, we can refine future synthesis pathways of HEOs with more precise tunability.

## 2. Materials and Methods:

Composition $(MgNiCuCoZn)_{0.167}(MnCr)_{0.083}O$ (J14CrMn) was synthesized for this study based on the "standard" J14 parent HEO [2]. The powders were mixed using the following binary oxide components: MgO (Sigma Aldrich, 342793, 99% trace metals basis, -325 mesh), NiO (Sigma Aldrich, 399523, -325 mesh, 99%), CuO (Sigma Aldrich, 450812, 99.99% trace metal basis), CoO (Sigma Aldrich, 343153, -325 mesh), ZnO (Sigma Aldrich, 255750, 99.99% trace metal basis), $Mn_3O_4$ (Sigma Aldrich, 377473, 97%), and $Cr_2O_3$ (Alfa Aesar, 12285, 99% trace metals). The powders were massed accordingly for the desired stoichiometry, listed in Table S1, and mixed/milled using a Spex 8000M Mixer Mill (Metuchen, NJ) using three 5 mm yttrium-stabilized zirconia (YSZ) milling media for 1 hour.

After mixing, powder was uniaxially pressed into a 25 mm pellet at 5500 pounds per square inch (psi) for 30 s (Carver Platen 2, Wabash, IN). The pellet was sintered in air at 1000° C for 8 hours with a ramp rate of 4 °C/min using a SentroTech (Strongsville, OH) STT-1700° C tube furnace and air quenched. Resulting phases of the reactively sintered target were identified using X-ray diffraction and are shown in Figure S1.

$(MgCoNiCuZn)_{0.167}(CrMn)_{0.0835}O$ films were deposited on (001)-oriented MgO substrates by ablation from a 248 nm KrF laser fired at 5 Hz and 1.7 J cm$^{-2}$ with a spot size of ~ 0.05 cm$^2$ for a total of 10000 pulses. The substrates were pre-annealed at 950 °C in base vacuum for 20 mins prior to being held at 400 °C in a pressure series (1, 2, 3, 4, 5, 7.5, 10, 15, 25, 50 mTorr) of pure $O_2$ throughout the deposition before cooling at the same pressure. The substrate-target distance was 7 cm.

X-ray diffraction (XRD) was performed on the reactively sintered targets for phase identification using a PANalytical X'Pert Pro MPD (Almelo, Netherlands). Measurements were taken using a theta-theta goniometer equipped with diverging beam optics and an X'celerator detector. Scans ranged from 15° to 80° 2θ with a step size of 0.033 °/step and dwell time of 200 sec/step. Phase identification was performed using Highscore Plus [45]. Films were measured in 2θ-ω with a Rigaku Smartlab diffractometer (Tokyo, Japan), using a 1.54 Å Cu Kα source, a Ge 220 monochromator, and parallel beam geometry. A 5 mm slit was used, with a 0.1-degree step size; for narrow scans the speed was 1 °/min, while for broad scans the speed was 12 °/min.

Transmission electron microscopy (TEM) sample preparation was performed using the Thermo Fisher Helios 660 NanoLab Dual-Beam FIB. The FIB lamellae were thinned down with a 5 kV ion beam, and final cleaning was carried out with 2 kV and 1 kV ion beams to reduce Ga$^+$ ion damage. The Thermo Fisher Talos F200X2 was used to obtain Selected Area Electron Diffraction (SAED) patterns at 200 kV.

X-ray absorption fine structure spectroscopy (XAFS) was performed on beamline 10-BM [46] at the Advanced Photon Source, Argonne National Laboratory (Lemont, IL). The films were suspended in a nylon washer and wrapped in Kapton tape to be measured in fluorescence yield using a Si (111) monochromator, detuned 50% for harmonic rejection, and a Hitachi 4-element Vortex ME-4 SDD (Tokyo, Japan) detector. The beamline was calibrated using a standard set of metal foils. In addition, these foils remained in the beam for transmission measurements allowing scan-by-scan energy alignment and ensuring energy calibration for each sample. The K-edges for cobalt (7709 eV), manganese (6539 eV), and chromium (5989 eV),  were targeted for this study due to their well-known variability in oxidation states, and each sample was scanned on every absorption edge between 3 and 5 times. The samples were scanned up to k = 13 Å$^{-1}$. Data was processed and analyzed using the Demeter software package [47].

Analysis of the XANES data consisted of extracting information on valence state and coordination through comparison to spectral reference standards from literature [50], [53], [54]. The literature standards were compared to the respective metal foils measured with our samples to ensure a correct energy grid alignment. Arcon, et al. demonstrated that the energy shift of the Cr K-edge is linearly proportional to oxidation state for materials systems containing the same ligand type[48]. By similar methods, estimation of the Co, Mn, and Cr oxidation states in J14CrMn was performed. The energy shift, $\Delta E_0$, was calculated by subtracting the metal edge energy from the edge energy for each sample. The edge energy, $E_0$, can be defined several ways including using white line maxima or peaks in the first derivative of the edge region. If consistency is maintained throughout each absorber, the relative shift between edge energies can be used to estimate average valence. $E_0$ for Cr and Co absorbers was chosen to be the middle peak in the first derivative, excluding the pre-edge, while the Mn $E_0$ was chosen to be the last peak in the first derivative due to noise. Coordination information is extracted through fingerprinting the features of the spectra standards to the collected data.

## 3. Results and Discussion:

XRD of the reactively sintered target contains three crystallographic phases including rocksalt, spinel, and tenorite, shown in Figure S1. Due to the similar scattering cross sections [49] of the constituent elements, we are unable to discern exact composition of each particular phase. Using PLD for film growth enables target stoichiometry to be controlled in the synthesis of single-phase epitaxial structures [22], [25], [50], [51]. Approximately 90 nm thick films grown epitaxially along the [001] direction at constant temperature on [100] MgO substrates with increasing pO$_2$ from 1 to 50 mTorr, shown in Figure S2. Crystallization appears at 2 mTorr pO$_2$. In the lower pressure regime, between 2 and 10 mTorr, the (002) film peak further crystalizes as evidenced by the formation of Pendellösung fringing, then begins to shift toward a smaller out-of-plane lattice parameter at 5 mTorr. Above 10 mTorr, the film peak appears to suggest strain, as evidenced by the asymmetry of the film peak and ultimately what may be the onset of phase separation. Full range scans from 30 – 100 °2θ shown in Figure S3 confirm epitaxial growth with no additional peaks present.

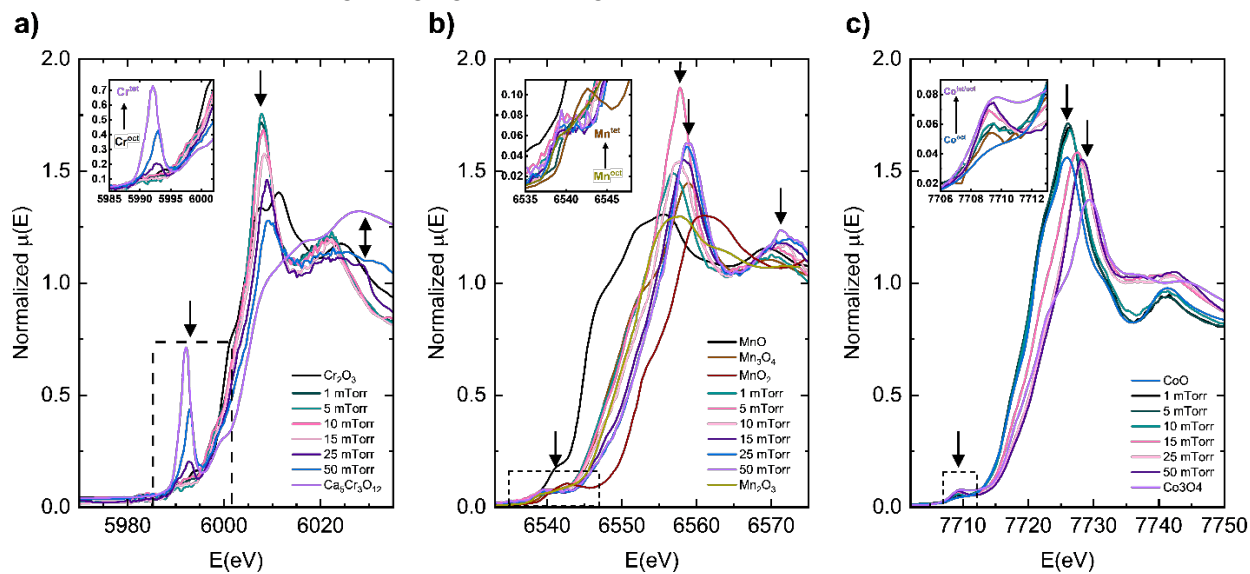

**Figure 1: X-ray near edge structure (XANES) of select absorbers a) Cr, b) Mn, and c) Co. Arrows indicate main features in spectra. Inset for each panel shows the pre-edge feature comparison between samples and literature references discussed in main text.**

The analysis of the XANES region can be used to provide information such as valence state, coordination number, and differences in local environment [52]. A comparison of measured XANES to known literature references [48], [53], [54] is shown in Figure 1. Changes in key fine structure features, denoted by arrows, show the local stereochemical evolution of absorbing cations. The pre-edge features in first row transition metals correspond to the 1s → 3d transition [55] and provide insight into coordination numbers. Initially, as prepared under the lower $pO_2$, we see a primarily $Cr^{3+}$ dominated environment shown in shown in Figure 1a. This is observed in the XANES data by a broad, pre-edge shoulder matching literature references for $Cr^{3+}$ in an octahedral environment [48]. As $pO_2$ is increased in subsequent depositions there is a systematic change in the Cr coordination indicative of 4-coordinated polyhedra [56], where a sharp pre-edge feature near 5992 eV develops and eventually comes to dominate the pre-edge structure. Further, the white line peak at ~ 6007 eV takes on a single peak instead of the double peak we see in the $3^+$ state and the region just above the edge (~ 6030 eV) increases in intensity toward that of the $6^+$ reference. Mn, shown in Figure 1b, increases in average valence between $3^+$ and $4^+$ with increasing $pO_2$ as indicated by the energy shift of the K-edge. The pre-edge features denoted by the arrow at ~ 6540 eV appear consistently throughout the pressure series and consistent with octahedral coordination. For Co, in Figure 1c, the spectra shift more towards the $Co^{3+}$ standard as pressure increases. The pre-edge feature also begins to increase intensity indicating a transition from octahedral toward a mixed occupation of octahedral/tetrahedral environment.

Using the strategy of Alcorn, et al.[48] mentioned previously, the relative change of the $\Delta E_0$, or the edge energy shift, vs the valence is plotted in Figure 2. The valence state can be estimated by the energy shift of the absorption edge relative to a known standard, typically a pure metal. The $\Delta E_0$ of the 5, 10, 25, and 50 mTorr samples were compared to standards and used to estimate valence in all three absorbers. Linear regression analysis of the $\Delta E_0$ versus valence of known standards provided a calibration line. Inserting the measured $\Delta E$ values for each sample into the corresponding linear equation enables one to determine the total average valence of the absorbing species. In Figure 2a, Cr appears to transition from an average of $4^+$ at 5 mTorr toward $5.5^+$ at 50 mTorr $pO_2$. Mn (Figure 2b) has an average valence trending from ~$3.5^+$ to $4^+$. Finally,

Co has an average valence of ~2.3 and ~2.6 at 5 mTorr and 50 mTorr pO$_2$, respectively, as shown in Figure 2c.

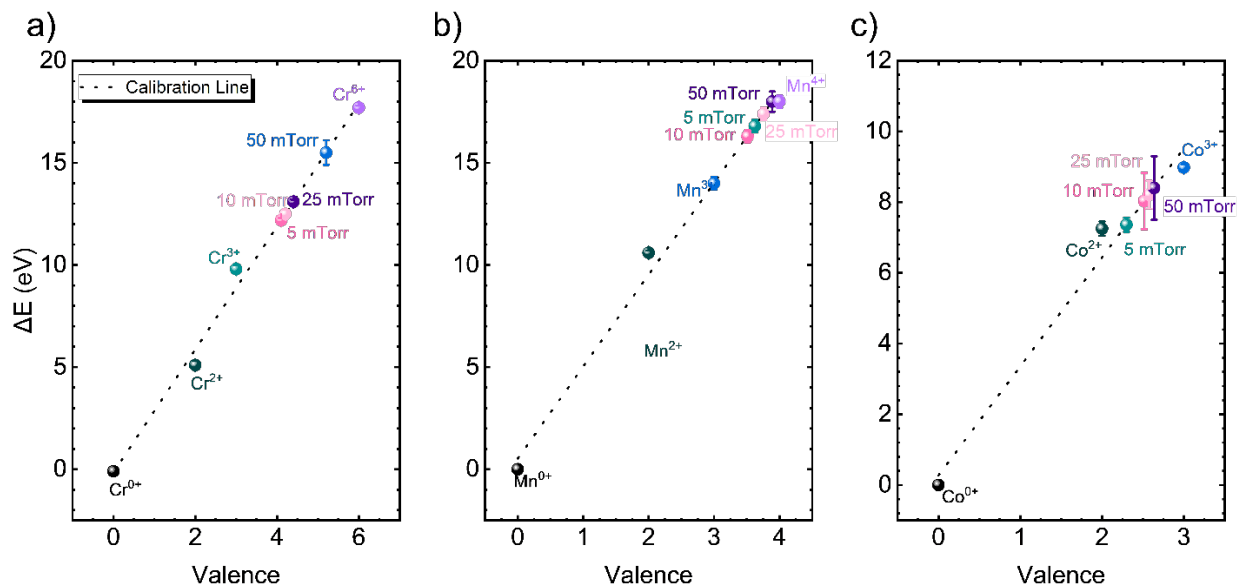

**Figure 2: Plot of ΔE vs. cation valence for a) Cr, b) Mn, and c) Co.**

Linear combination fitting (LCF) was performed to support the oxidation trend results across Co, Cr, and Mn absorbers, respectively, as shown in Figures 3 a-c. The Cr absorber data begins at the low pressures with a mixed oxidation heavily weighted toward 3$^+$. Above 15 mTorr, the fraction of 3$^+$ begins to decrease, with an appearance of 6$^+$ beginning at 25 mTorr. At 50 mTorr, there is a mixed oxidation of 3$^+$ and 6$^+$ with fractions of 0.53 and 0.43 respectively, the remaining fraction being most likely held by 2$^+$. The Mn absorber starts fully in the 3$^+$ oxidation state for lower pressures, with 4$^+$ oxidation state emerging above 7.5 mTorr deposition pressure. At 15 mTorr, 4$^+$ overtakes 3$^+$ as the predominant state with fractions of 0.57 to 0.43. The fraction of 4$^+$ increases to ~ 0.7 at the 25 and 50 mTorr data points. Finally, the Co absorber has an oxidation of 2$^+$ from 1-10 mTorr pO$_2$, a mixed oxidation with majority 2$^+$ at 15 mTorr, and then mixed with a majority 3$^+$ at higher pO$_2$.

It is important to note the limitations of conventional LCF, especially within the context of HEOs containing several transition metal elements. The challenge is two-fold: finding a literature or measurable standard to use as a component and accounting for relative changes in feature intensity (like white lines) due to sample geometry and beam polarization effects. Here, we used LCF to ascertain valence trends in Cr, Mn, and Co absorbers by fitting a weighted sum of standards found in the literature based on valence state. It is near impossible to account for all deviations in white line intensities or pre-edge features given that not all permutations of valence, geometry, and coordination have specific reference standards or known crystal systems. The takeaway of our multifaceted XANES analysis is that the trends are consistent and generally in agreement, thus strengthening confidence in our findings.

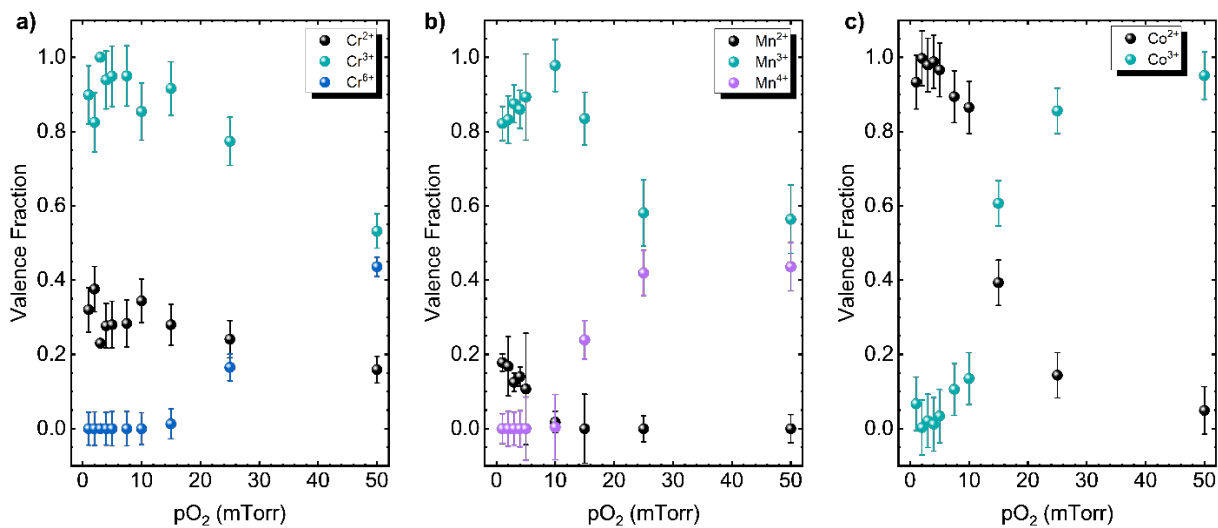

*Figure 3: Plot of Valence Fraction vs pO₂ achieved by LCF. Example of fitting in Figure S4.*

Figure 4 illustrates the magnitude and imaginary parts of the Fourier transformed k2-weighted EXAFS for a number of cases where we see a clear and observable trend within the first coordination shell. To investigate the first coordination shell, we fitted the spectra using a theoretical model with a single scattering path between the absorber and nearest neighbor oxygens. At low pressures, the film appears to be amorphous per XRD results, shown in Figure

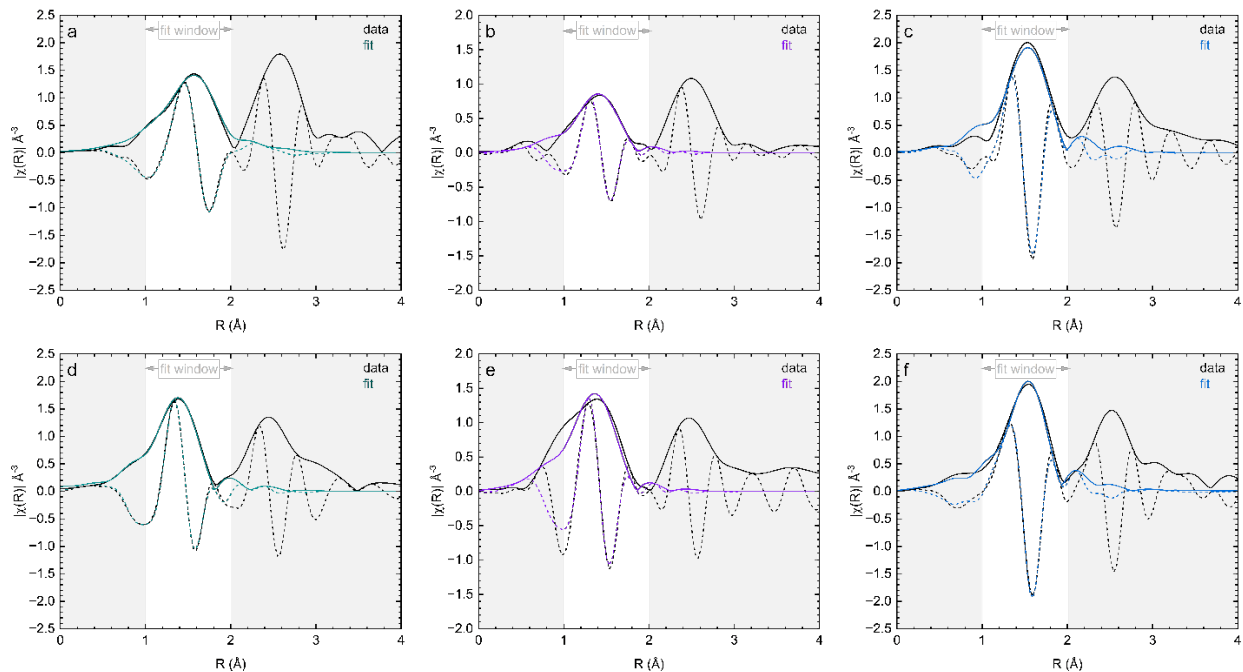

***Figure 4: Magnitude and imaginary parts of the Fourier transformed $k^2$-weighted EXAFS for a) Co at 3 mTorr $O_2$, b) Mn at 1 mTorr $O_2$, c) Cr at 2 mTorr $O_2$, d) Co at 25 mTorr $O_2$, e) Mn at 15 mTorr $O_2$, f) Cr at 15 mTorr $O_2$.***

S2. EXAFS, however, does not depend on long-range order in the system. Therefore, to understand the synthesis trends from a local perspective we choose the 1 mTorr sample as a baseline for the first shell environment. For some atoms, like Mn where the coordination number reaches a steady state with increasing pressure, we can extrapolate radial distances and unit cell parameters. These results match the bond lengths and unit cell parameters found via XRD for the samples, thus giving a reasonable basis for studying the metal-oxygen bond. The second shell corresponds to the metal-metal scattering paths, which remain largely stable. Modeling of this length, however, is limited by the stoichiometric ratios and the effects of various scatterers similar in mass at similar distances. The range from $k = 6 - 8$ Å$^{-1}$ demonstrates a poor fit for first shell scatterers. This is because this range is largely indicative of the heavier scatterers in the sample. A linear combination of both fit sets yields an approximate solution for the full structure. Note that although a solution exists, the uncertainties arising from the unknown mass of the metal-metal scatterer restrict the amount of information that can be confidently drawn. Therefore, it is better used as a guide rather than an exact fit.

The results for Mn display a relatively stable coordination environment, within uncertainty, in the $pO_2$ range corresponding to highest crystallinity. While there are minor fluctuations, likely due to the noise of the Mn spectra, the short range of the fit restricts the number of independent variables. Our conclusions are drawn from the first shell fits for each absorber element. Accurate modeling of the metal-metal distance would require a variable set of metals with closely similar atomic masses. We trade the information and resolution of the results for more clarity on the well-known bonding between metal and oxygen. Therefore, when we consider the results shown in Table S2. For Mn, the coordination number is 4.5 +/- ~0.5. This indicates a nearly 50/50 mix of Mn in both the 4-fold and 6-fold coordinated states at 10mTorr. For Cr, the coordination environment undergoes an onset phase-transition-like process where the coordination number starts at 6, but then converts to 4-fold symmetry after about 15 mTorr of $pO_2$ is introduced. We theorize that once there is sufficient oxygen for cations to undergo a change of valence, and subsequent change in ionic radius, a more energetically favorable polyhedral state, based on close packing, is formed around certain cations that becomes incorporated into the larger structure. We propose that in-plane bonding remains strong; this is the 4 coordinated, tetragonal state environment we see dominating at higher $pO_2$. The out-of-plane bonding, however, exhibits compression and expansion based on the partial pressure of $O_2$. A handful of atoms, like Mn and Cu, hold the larger superstructure together, as demonstrated by the unchanged mixed-fold coordination we would expect. This suggests that a spinel might be forming as the bonds are strained with the increased oxygen sites and decreased valence electrons available to form bonds.

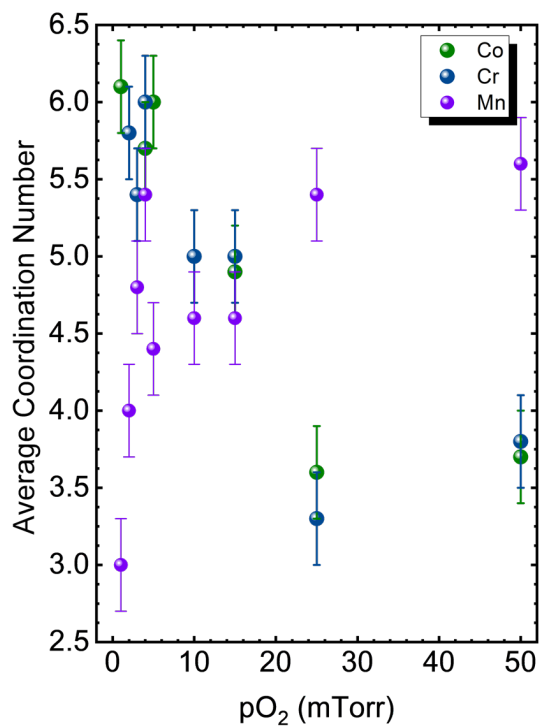

***Figure 5: Coordination number vs. oxygen partial pressure for Co, Cr, and Mn absorbers.***

Selected Area Electron Diffraction (SAED) experiments using transmission electron microscopy (TEM) were carried out to investigate the structural nuances in the thin films grown at 5 mTorr and 50 mTorr $O_2$. From the TEM diffraction patterns shown in Figure 6, we observed a clear variation in the crystal structural of the thin films grown at different $O_2$ partial pressures from the selected area electron diffraction experiment. J14CrMn grown at 5 mTorr indicates a rock salt crystal structure, while J14CrMn grown at 25 mTorr and 50 mTorr has rock salt structure with additional reflections (marked by green boxes in Figure 6). To further understand the origin of the extra reflections, selected area electron diffraction was performed along [110] zone axis for 50mTorr sample (shown in Figure S5 in supplementary). These extra reflections indicate local structural variation that are consistent with spinel secondary phase observed in XRD. This is also associated with the changes in the oxidation state of cations shown in Figure 2. Further TEM studies are being carried out to probe the local variation in oxidation states.

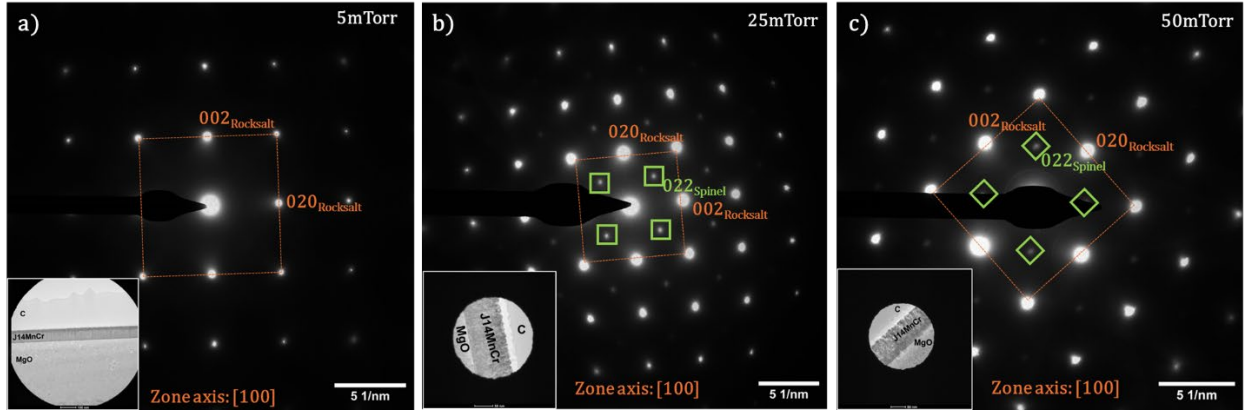

*Figure 6: Selected Area Electron Diffraction of J14MnCr at a) 5 mTorr O2 b) 25 mTorr O2 c) 50 mTorr O2. Green boxes in b) & c) highlight the presence of extra reflections that are typically absent in perfect rock salt structure.*

Nonequilibrium kinetics by pulsed laser deposition (PLD) is uniquely suited to enable cation's charge state control in high entropy oxides. This is achieved through the metastabilization of unique macrostates at substrate temperatures lower than the entropy stabilization temperature which provides a rich landscape of local structures and oxidation states [25], [23], [57], [58]. For instance, Kotsonis et al. noted a sharp change in the out-of-plane lattice parameter of J14 thin films with substrate temperature. Electron energy loss spectroscopy (EELS) as well as x-ray absorption spectroscopy (XAS) were employed to inspect the valence state of the core elements in the thin films [23]. Local analysis reveals that at relatively low substrate temperatures, a significant Co $3^+$ concentration is present and is attributed to cation vacancies.

We believe that the high configurational entropy and metastability inherent to HEOs facilitate hosting more accepting local environments coupled with unusual coordination and defect chemistry. As such, the degree of tunability and range of lattice parameter and volume changes should increase with compositional complexity. To highlight this, Figure 7 shows two sets of films of $Mg_{0.83}Cr_{0.17}O$ grown 5 and 50 mTorr $pO_2$ at temperatures of 300 and 500 °C, respectively. While the films display a decrease in the out of plane lattice parameter with increasing deposition pressure, the degree of the shift is less pronounced than the high entropy case. Therefore, we need specialized characterization techniques to understand the local structures and cation valence in high configurational entropy systems and how they deviate from their lower entropy counterparts.

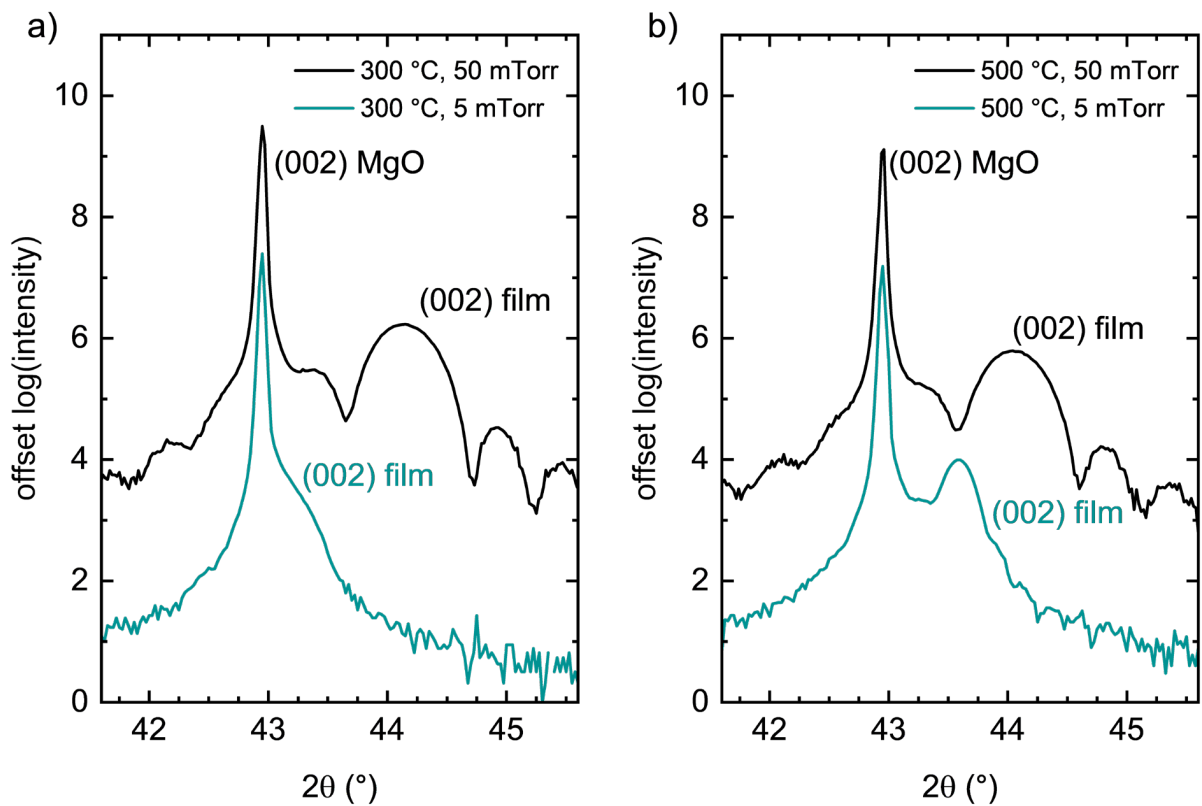

*Figure 7: XRD diffraction of two $Mg_{0.83}Cr_{0.17}O$ films grown at a) 300 °C, 5 and 50 mTorr $pO_2$, and b) 500 °C, 5 and 50 mTorr $pO_2$ at a constant 1.2 $J•cm^{-2}$ fluence.*

By examining the behavior of local structure as a function of deposition pressure, we can better understand the long-range behavior of these materials. Combining the information from XANES and EXAFS gives us a more complete picture of how individual atomic species respond to the evolving structure as deposition pressure is varied. While the transition between oxidation states can be modeled with a power law, it is important to consider that it may not be possible to tune with a great deal of resolution. The onset occurs over a very short range of partial pressures. Maintaining specific levels during deposition is an ongoing area of study to control these tunable aspects. Notably, a localized expansion was found around the Mn atom. This indicates that not all atoms are equally affected in their transition. These differences could be ascribed to the relative masses of other cations. Soft x-ray sources might yield information about the oxygen environment; however, due to the expanse of bonding options deconvoluting the oxygen, XANES is speculative at best.

One of the motivations driving the development and study of HEOs is their potential properties. The seminal structure HEO J14 is found to have long-range antiferromagnetic ground state with unique spin-state interactions [59]. Testing the magnetic properties as a function of deposition pressure of J14CrMn may allow us to further understand the effects of local structure on global properties. Future research could delve into the changes in magnetic structure associated with different oxidation states. Furthermore, if experimental findings could be corroborated with theoretical calculations, then this allows us to finely tune the structure and deposition conditions for HEOs.

## 4. Conclusion:

Seven-component HEO $(MgNiCuCoZn)_{0.167}(MnCr)_{0.083}O$ (J14CrMn) was successfully synthesized using pulsed laser deposition with varying initial conditions. This study examined the local structure of this material to gain a deeper understanding of how this emerging class of materials behaves. For J14CrMn, the lattice parameter decreases as partial oxygen pressure increases, which is the opposite trend of the seminal J14 structure. According to XANES analysis, the oxidation state and coordination number are also dependent on the partial pressure of oxygen, with a phase shift occurring as pressure is increased in the Cr, Mn, and Co absorbers. A close study of the metal-oxygen paths via EXAFS supports the XANES data, as Cr shifts from 6-fold to energetically favorable 4-fold when sufficient oxygen is introduced to the system. These changes in coordination suggest a spinel phase formation as a result of bond strains introduced by fewer valence sites. Finally, SAED results indicate the spinel structure forming as deposition pressure is increased, thus supporting both the XRD and EXAFS analysis. From these results, we theorize that while in-plane bonding is stable, it is the out-of-plane bond lengths that change with pressure. We observe a select few atoms remaining consistent and dictate the larger superstructure, while other atoms shift in coordination.

**Acknowledgements:**

The authors acknowledge the support from NSF through the Materials Research Science and Engineering Center DMR 2011839. Sector 10 MRCAT operations are supported by the Department of Energy and the MRCAT member institutions. This research used resources of the Advanced Photon Source, a US Department of Energy (DOE) Office of Science User Facility operated for the DOE Office of Science by Argonne National Laboratory under Contract No. DE-AC02-06CH11357.

**Declarations of Competing Interest:**

The authors declare that they have no known competing financial interests or personal relationships that could have appeared to influence the work reported in this paper.

**Contributions:**

C.M.R. conceived and supervised the project. Targets were synthesized by G.E.N. and thin films were grown by M.W. and S.S.I.A., respectively. XAFS measurement and analysis were performed by G.R.B., J.P.B., J.T.W., G.E.N., and C.M.R. TEM measurement and analysis was performed by S.V.G.A. and N.A. This paper was written by G.E.N., G.R.B., J.P.B., and C.M.R. with input from all authors.

**Supplemental Information:**

*Table S1: Stoichiometric mass amounts of metal oxide powders used to make J14MnCr.*

| Oxide | Amount (g) |
|-------|-----------|
| ZnO | 1.620 |
| MgO | 0.807 |
| CoO | 1.497 |
| CuO | 1.588 |
| NiO | 1.480 |
| Mn3O4 | 0.770 |
| Cr2O3 | 0.390 |

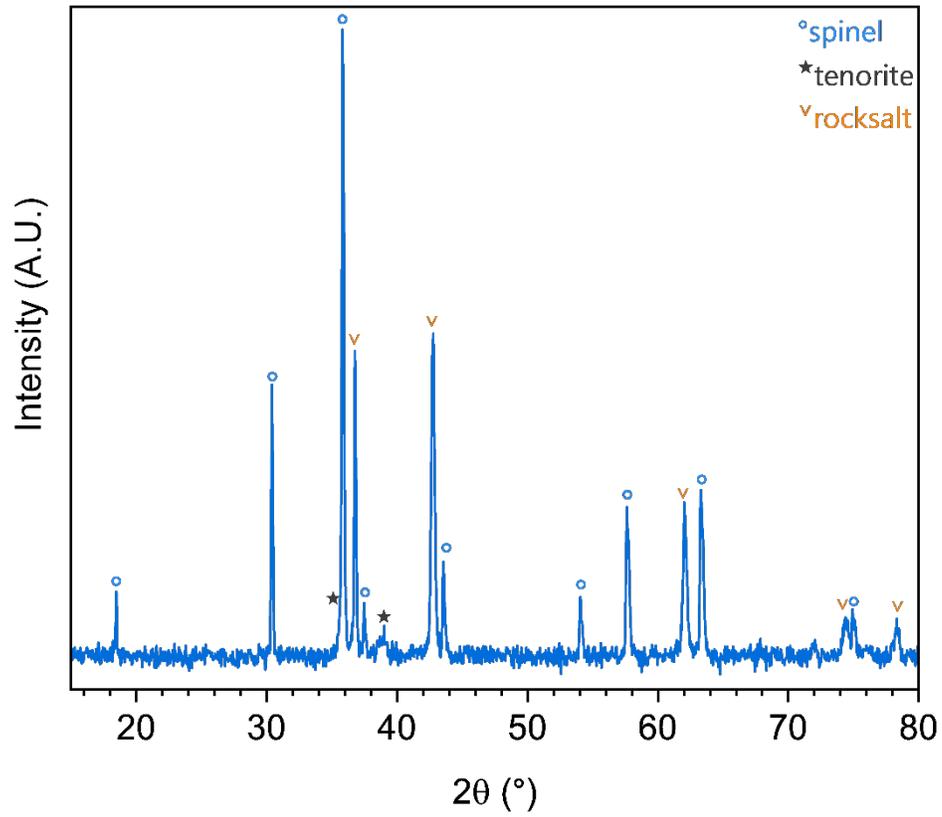

*Figure S8: X-ray diffractograms of reactively sintered target J14CrMn. Each target exhibits three crystallographic phases: spinel (Fd-3m), rocksalt (Fm-3m), and tenorite (C2/c).*

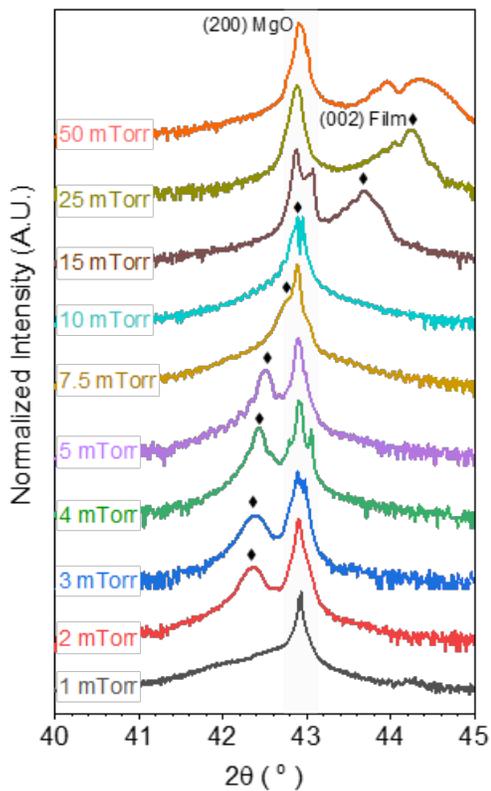

*Figure S9: X-ray diffractograms centered on the (200) reflection of the MgO substrate depicting the contraction in lattice parameter of the thin films as pressure increases with the beginnings of phase separation behavior above 10 mTorr.*

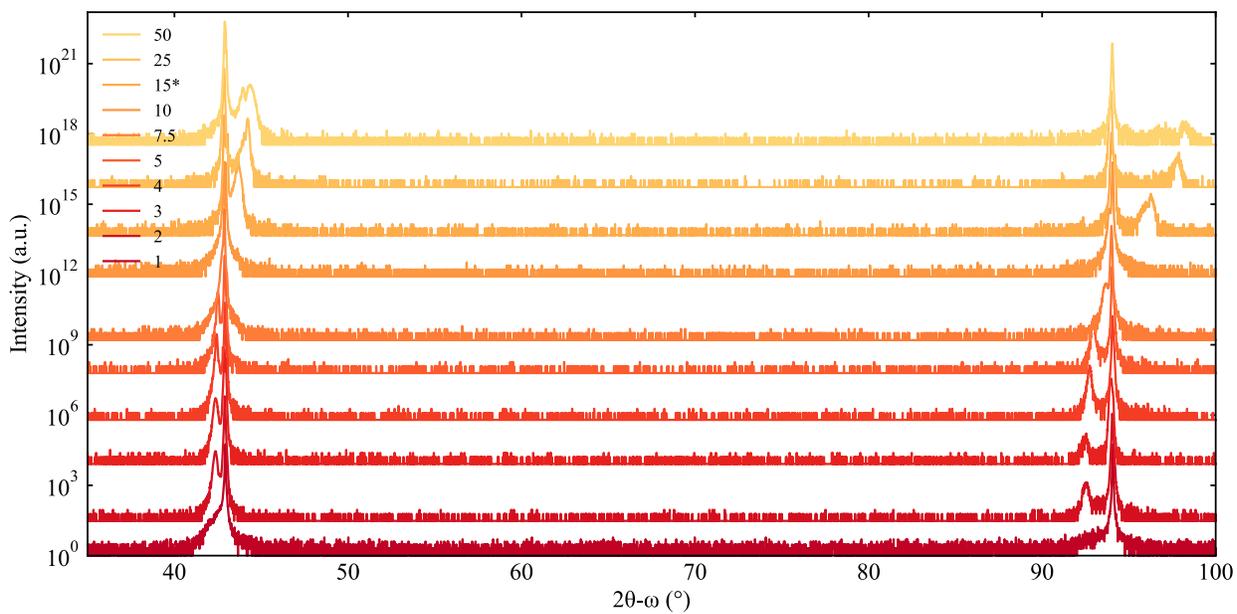

*Figure S10: Long range X-ray diffractograms of thin films grown on MgO, indicating epitaxial growth as no additional peaks appear.*

**Linear Combination Fitting:**

Linear combination fitting (LCF) is a method that can be used to give an approximate quantification of oxidation states. Using a combination of sample spectra, LCF attempts to reconstruct the sample spectrum and gives a report detailing the percentage that each sample has contributed to the fit. LCF also reports the R-factor and the reduced chi squared. The accuracy of LCF relies heavily on how the standards represent the components that are in the sample. For best results, the sample spectra should be normalized and aligned before starting LCF. LCF was performed using Athena [1]. First, standards were imported, properly calibrated, and rebinned. For each fit, except the Cr fit, the options of 'all weights between 0 and 1' and 'force weights to sum to 1' were selected. Data was fit in both normalized and derivative space. For Cr, the 'force weights to sum to 1' caused errors in the fit, so the option to 'force the sum of the weights to 1' was not selected. The errors were most likely due to a difference in normalization between the standards from the literature and the measured data. The fit window was adjusted so that only relevant data would be considered, for example a post-edge peak in the data that is not matched in the standards could be disregarded due to the sample being a complex oxide. For consistency, standards were marked and 'use marked groups' was selected before each fit. Each LCF was run twice to check for consistency in the fits. After the data was validated, the reports were saved as column data.

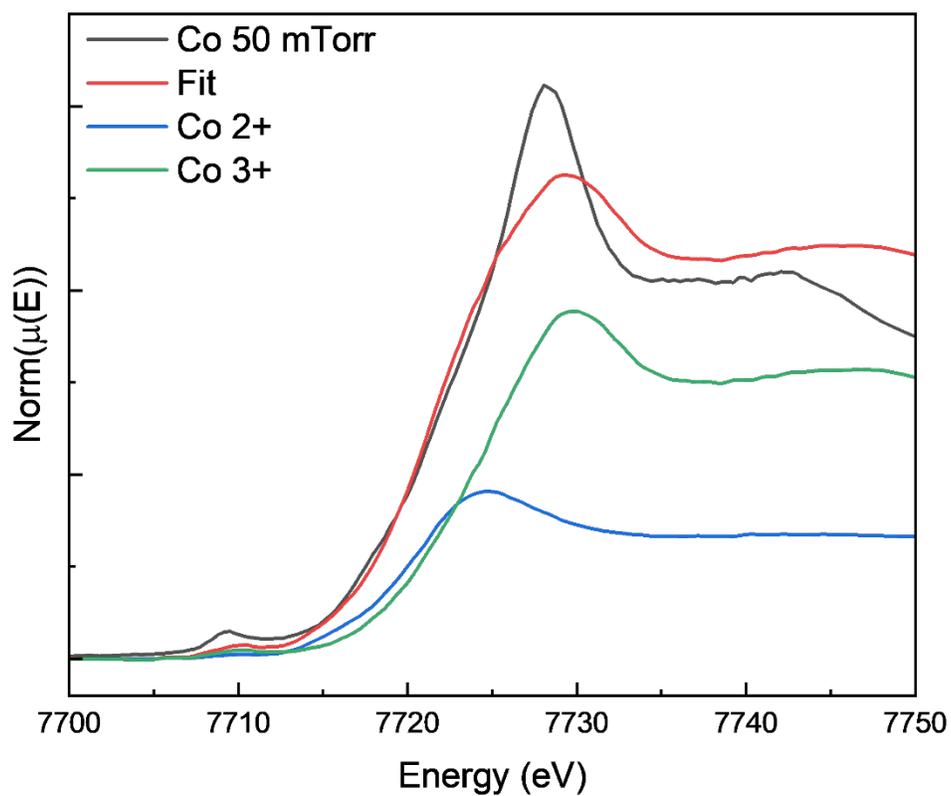

*Figure S11: Example LCF fit of Co 2+ and Co 3+ Standards on the Co K-edge of 50 mTorr sample.*

**Table S2: Best fit average results for first coordination shell of Cr, Mn, and Co, respectively.**

| | Cr | | | Mn | | | Co | | |
|---|---|---|---|---|---|---|---|---|---|
| pO₂ (mTorr) | N | R (Å) | σ² (Å²) | N | R (Å) | σ² (Å²) | N | R (Å) | σ² (Å²) |
| 1 | | 2.08(1) | 0.002(1) | 2.9(3) | 1.90(1) | 0.007(1) | 6.0(3) | 2.10(1) | 0.012(1) |
| 2 | 5.8(3) | 2.09(1) | 0.006(1) | 3.9(3) | 1.88(1) | 0.001(1) | 6.7(3) | 2.09(1) | 0.013(1) |
| 3 | 5.4(3) | 2.00(1) | 0.001(1) | 4.7(3) | 1.87(1) | 0.007(1) | 5.3(3) | 2.09(1) | 0.010(1) |
| 4 | 5.9(3) | 2.09(1) | 0.002(1) | 5.3(3) | 1.88(1) | 0.011(1) | 5.6(3) | 2.10(1) | 0.010(1) |
| 5 | | | | 4.3(3) | 1.90(1) | 0.008(1) | 6.0(3) | 2.08(1) | 0.010(1) |
| 7.5 | 6.4(3) | 2.10(1) | 0.002(1) | 4.2(3) | 1.86(1) | 0.002(1) | 5.2(3) | 2.09(1) | 0.012(1) |
| 10 | 5.0(3) | 2.10(1) | 0.003(1) | 4.5(3) | 1.88(1) | 0.001(1) | 4.9(3) | 1.95(1) | 0.011(1) |
| 15 | 5.0(3) | 2.09(1) | 0.001(1) | 4.6(3) | 1.87(1) | 0.002(1) | 3.6(3) | 1.90(1) | 0.002(1) |
| 25 | 3.3(3) | 2.11(1) | 0.001(1) | 5.4(3) | 1.88(1) | 0.002(1) | 3.6(3) | 1.90(1) | 0.001(1) |
| 50 | 3.7(3) | 2.19(1) | 0.002(1) | 5.5(3) | 1.88(1) | 0.002(1) | 6.0(3) | 2.10(1) | 0.012(1) |

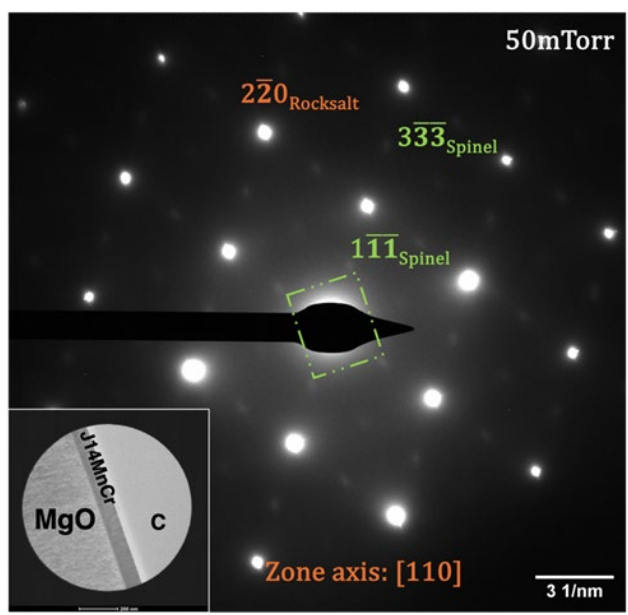

***Figure S12:*** *Selected Area Electron Diffraction of J14MnCr at 50 mTorr O2 [110] zone axis. Green boxes highlight the presence of extra reflections that are typically absent in perfect rocksalt structure.*

**Supplemental References:**